# Electromagnetic fields, size, and copy of a single photon

Shan-Liang Liu

Shandong Key Laboratory of Optical Communication Science and Technology, School of Physical Science
and Information Engineering, Liaocheng University, Shandong 252059, People's Republic of China

Photons are almost involved in each field of science and daily life of everyone. However, there are still some fundamental and puzzling questions such as what a photon is. The expressions of electromagnetic fields of a photon are here proposed. On the basis of the present expressions, we calculate the energy, momentum, and spin angular momentum of a photon, derive the relations between the photon size and wavelength, and reveal the differences between a photon and its copy. The results show that the present expressions properly describe the particle characteristics of a photon; the length of a photon is half of the wavelength, and the radius is proportional to square root of the wavelength; a photon can ionize a hydrogen atom at the ground state only if its radius is less than the Bohr radius; a photon and its copy have the phase difference of $\pi$ and constitute a phase-entangled photon pair; the phase-entangled n-photon train results from the sequential stimulated emissions and belongs to the Fock state. A laser beam is an ensemble of the n-photon trains and belongs to the coherent state. The threshold power of a laser is equal to the power of the n-photon train. These provide a bridge between the wave theory of light and quantum optics and will further advance research and application of the related fields.

## I. INTRODUCTION

Light is almost involved in each field of science and daily life of everyone, and yet light's true nature has eluded us for centuries. Albert Einstein successfully explained the photoelectric effect in 1905 by the assumption that light is composed of photons and showed that a photon has constant energy $h\nu$ and momentum $h/\lambda$ where h is the Planck constant, $\nu$ is the frequency, and $\lambda$ is the wavelength. The theory of relativity tell us that a photon has zero rest mass and always moves at the speed of $c=\lambda\nu$ in vacuum. However, these answers are not satisfactory. Roy J. Glauber once jokingly summarized his theory of photo detection by the sentence: "I don't know anything about photons, but I know one when I see one" [1]. Single photons are essential for the fundamental study of the quantum mechanics [2,3] and the development of photonic quantum technologies such as optical quantum computing [4-6] and quantum communication [7-9]. However, there is not still a satisfactory answer to the problem what a photon is [10].

The wave-like properties of microscopic particles such as an electron and proton are usually described by wave functions in quantum mechanics, and the particle-like behaviors are governed by the classical or relativistic dynamic equations where they are usually simplified as a point with mass. Size of an atom or ion is usually represented by the orbit diameter of the outer electron in it, while radius of the atomic nucleus is proportional to cubic root of the mass number. The experiments have indicated that a single photon can locate in very small space [11,12] and very short time duration [13,14], but how to know size of a photon is still a puzzling question. The wave-like properties of light or photons are well described by the classical theory of electromagnetic fields. A monochromatic plane wave indicates that a single photon would occur at any place with the same probability, which clearly contradicts the well-known experimental fact that the photons out of a single-photon source can be detected only at right time by the detector at right position, and so it is unable to properly describe a photon. A wave packet results from superposition of plane waves with different wave vectors in the classical theory of electromagnetic fields and is obviously incomparable with the well-known fact that a photon has the definite energy and momentum, and so it is unable to properly describe a photon either. How to properly describe the electromagnetic fields of a single photon is still a fundamental and unresolved question in physics. The no-cloning theorem, which says that one cannot make a perfect cloning of an unknown quantum state [15], is considered to be heart of security of quantum cryptography [16,17]. The stimulated emission is natural way of cloning a photon [18]. However, any difference between a photon and its copy has not been found.

The expressions of electromagnetic fields of a photon are here proposed. On the basis of the present expressions and the known experimental facts, we calculate the energy, momentum, and spin angular momentum of a photon, derive the relations between the photon size and wavelength, and reveal the differences between a photon and its copy. These provide a bridge between the wave theory of light and quantum optics and will further advance the research and application of the related fields.

## II. ELECTROMAGNETIC FIELDS
## OF A PHOTON

Since light is composed of photons and the Maxwell equations are linear in vacuum, behaviors of a photon as well as light are governed by the Maxwell's equations, and the electromagnetic fields of a photon can be described by the special solutions of the Maxwell equations

$$\boldsymbol{E}_1 = f_1 E_0 Re(e^{i\varphi}\boldsymbol{u}), \qquad (1)$$
$$\boldsymbol{B}_1 = \boldsymbol{e}_z \times \boldsymbol{E}_1/c, \qquad (2)$$
$$f_1 = \begin{cases} 1, \rho < \rho_0 \ and \ \varphi \in [\varphi_1,\varphi_0] \\ 0, \rho > \rho_0 \ or \ \varphi \notin [\varphi_1,\varphi_0] \end{cases}. \qquad (3)$$

where $\varphi = kz - \omega t$, $\omega = 2\pi\nu$, $k = 2\pi/\lambda$, $\boldsymbol{u} = \boldsymbol{e}_x\cos\alpha + \boldsymbol{e}_y\sin\alpha e^{i\delta}$, $\alpha$ and $\delta$ are the polarization parameters, $E_0$ is a real constant, and $f_1$ is a profile function which describes a cylinder with the length $z_0 - z_1 = \lambda(\varphi_0 - \varphi_1)/2\pi$ and radius $\rho_0$. The photon position varies from $-\infty$ to $\infty$ with time $t$ along the motion direction $\boldsymbol{e}_z$. For a linearly-polarized photon $\delta = 0$ or $\pi$, the electric field varies with phase $\varphi$ in magnitude and has the maximum $E_0$. For a circularly-polarized photon $\alpha = \pi/4$ and $\delta = \pm\pi/2$, the electric field has the constant magnitude of $E_0\sin(\pi/4)$, and its direction varies with the phase. An elliptically-polarized photon corresponds to the other values of $\alpha$ and $\delta$, and both magnitude and direction of the electric field vary with the phase.

## III. SIZE OF A PHOTON

There are no electromagnetic fields in the region of $\rho > \rho_0$. The energy density $\varepsilon_0 E_1^2$ in a photon is independent of $\rho$ where $\varepsilon_0$ is the vacuum permittivity, but it depends on the polarization state. The photon energy is given by the integral

$$h\nu = \int_{z_1}^{z_0} \varepsilon_0 E_1^2 \pi \rho_0^2 dz. \qquad (4)$$

### A. Length of a photon.

Integrating the right side of Eq. (4) gives

$$h\nu = \varepsilon_0 E_0^2 \rho_0^2 \lambda(\Delta_- + C\sin\Delta_-)/4, \qquad (5)$$

where $C = cos^2\alpha \cos\Delta_+ + sin^2\alpha \sin(\Delta_+ + 2\delta)$ and $\Delta_\pm = \varphi_0 \pm \varphi_1$. It is well known that energy of a photon is independent of the polarization state, i.e. the parameters $\alpha$ and $\delta$ should not be included in Eq. (5). This requires $\sin\Delta_- = 0$, i.e. $\Delta_-$ be a multiple of $\pi$. Since a photon is the most fundamental quantum of electromagnetic fields, $\Delta_-$ must take the minimum $\pi$, i.e. $\varphi_1 = \varphi_0 - \pi$, and Eq. (5) reduces to

$$h\nu = \varepsilon_0 E_0^2 \rho_0^2 \lambda/4, \qquad (6)$$

which indicates that the radius $\rho_0$ as well as $E_0$ is independent of the polarization state.

A photon has the momentum density $\boldsymbol{e}_z \varepsilon_0 E_1^2/c$, and its momentum is given by the integral

$$\frac{\boldsymbol{e}_z}{c} \int_{z_1}^{z_0} \varepsilon_0 E_1^2 \pi \rho_0^2 dz = \frac{h\boldsymbol{e}_z}{\lambda}. \qquad (7)$$

The calculated result is the same as the known momentum expression of a photon. It follows that the phase interval of a photon is $\pi$, i.e. the longitudinal length of a photon is half of the wavelength $\lambda/2$ in spatial domain and half of the cycle $T/2$ in temporal domain where $T = 1/\nu$. The half-cycle pulses have been used to ionize the Rydberg atoms and found to be linear polarization and unipolar [19].This means that the electric fields in a linearly-polarized photon are in the same direction, and $\sin\varphi_0 = \pm 1$.

The spin angular momentum (SAM) density of free electromagnetic fields in propagation direction is given by $\varepsilon_0 \boldsymbol{E} \times \boldsymbol{A}$ where $\boldsymbol{A}$ is the vector potential [20]. SAM of a photon is given by the integral

$$\boldsymbol{L}_s = \int_{z_1}^{z_0} \varepsilon_0 \boldsymbol{E}_1 \times \boldsymbol{A}_1 \pi \rho_0^2 dz, \qquad (8)$$

where $\boldsymbol{A}_1$ is the vector potential of a photon. Using the relation $\boldsymbol{E}_1 = -\partial \boldsymbol{A}_1/\partial t$ and Eq. (1) gives the vector potential of a photon

$$\boldsymbol{A}_1 = -f_1 A_0 Re(ie^{i\varphi}\boldsymbol{u}), \qquad (9)$$

where $A_0 = E_0/\omega$. Substituting Eq. (9) into Eq. (8) gives

$$\boldsymbol{L}_s = -\hbar \sin(2\alpha) \sin\delta \ \boldsymbol{e}_z, \qquad (10)$$

where $\hbar = h/2\pi$ is the reduced Planck constant. For a circularly-polarized photon $\alpha = \pi/4$, $L_s = -\hbar$ at $\delta = \pi/2$, and $L_s = \hbar$ at $\delta = -\pi/2$. For a linearly-polarized photon $\delta = 0$ or $\pi$, $L_s = 0$. For an elliptically-polarized photon, $L_s$ varies with the polarization parameters $\alpha$ and $\delta$. Equation (10) is in good agreement with the known experimental results [21].

### B. Radius of a photon.

After the electron in a hydrogen atom meets a photon, the change rate of the electron momentum is governed by the dynamic equation

$$\frac{d\boldsymbol{p}_e}{dt} = \boldsymbol{f}_c - e\boldsymbol{E}_1 - e\boldsymbol{v} \times \boldsymbol{B}_1, \qquad (11)$$

where $\boldsymbol{f}_c$ is the Coulomb force acted on the electron, e is the elementary charge, and $\boldsymbol{v}$ is the velocity of the electron. Taking scalar product of Eq. (11) and $\boldsymbol{v}$d$t = d\boldsymbol{r}$ gives

$$\boldsymbol{v} \cdot d\boldsymbol{p}_e = \boldsymbol{f}_c \cdot d\boldsymbol{r} - e\boldsymbol{v} \cdot \boldsymbol{E}_1 dt. \qquad (12)$$

Since $v \ll c$ for a hydrogen atom, $\boldsymbol{p}_e = m_0\boldsymbol{v}$ where $m_0$ is the rest mass of an electron, and $\boldsymbol{v} \cdot d\boldsymbol{p}_e = d(\boldsymbol{p}_e \cdot \boldsymbol{p}_e)/2m_0$. Integrating Eq. (12) gives change of the kinetic energy in the ionization process

$$\frac{p_{ef}^2 - p_{e0}^2}{2m_0} = W - \frac{e}{m_0}\int_0^{T/2} \boldsymbol{p}_e \cdot \boldsymbol{E}_1 dt. \qquad (13)$$

where $W = -2U_i$ is work of the Coulomb force on the electron in the ionization process, $U_i = p_{e0}^2/2m_0$ is the ionization energy according to the Bohr's theory. The last term is the work of the electric field of the photon on the electron in the absorption process and has the maximum $h\nu$ if the photon is linearly polarized and $\boldsymbol{p}_e \cdot \boldsymbol{E}_1 = -p_e E_1$ in the absorption process. The photon energy required for ionization of $p_{ef} = 0$ has the minimum $U_i$, and Eq. (13) reduces to

$$-\frac{e}{m_0}\int_0^{T/2} \boldsymbol{p}_e \cdot \boldsymbol{E}_1 dt = U_i. \qquad (14)$$

Since v<<c in a hydrogen atom, $|v \times B_1|<<E_1$, the last term in Eq. (11) can be negligible, and

$$dp_e = f_c dt - eE_1 dt, \qquad (15)$$

The vector potential $A1T = -A10$ when the photon is completely absorbed at $t=T/2$ where $A10$ is the vector potential at $t=0$. Using the relation $E_1 = -\partial A_1/\partial t$ and integrating Eq. (15) gives

$$p_{eT} - p_{e0} = \int_0^{T/2} f_c dt - 2eA_{10}. \qquad (16)$$

Since $p_c \cdot E_1 = -p_c E_1$ and $p_c \cdot f_c = 0$ in a hydrogen atom, $f_c \cdot E_1 = 0$ and $f_c \cdot A_1 = 0$. Integrating Eq. (14) gives by use of Eq. (16)

$$2e^2 A_0^2 + 2ep_{e0} A_0 = m_0 U_i. \qquad (17)$$

Solving Eq. (17) gives

$$A_0 = (\sqrt{2} - 1)\sqrt{2m_0 h\nu/2}/e. \qquad (18)$$

Using $E_0 = \omega A_0$ gives amplitude of the electric field

$$E_0 = (\sqrt{2} - 1)\pi\nu\sqrt{2m_0 h\nu}/e. \qquad (19)$$

Substituting Eq. (19) into Eq. (6) gives the photon radius

$$\rho_0 = \frac{2\sqrt{2r_e\lambda}}{(\sqrt{2}-1)\pi}, r_e = \frac{e^2}{4\pi\varepsilon_0 m_0 c^2}, \qquad (20)$$

where $r_e$ is the classical radius of an electron. Since the length of a photon is equal to half of the wavelength and the radius is proportional to square root of the wavelength, the size and shape of a photon vary with the photon energy or wavelength. A photon is in shape like a thin stick if its energy is lower than the rest energy of an electron and like a plate if its radius is smaller than the classical radius of an electron. For a photon of $h\nu=13.6$ *eV*, the photon radius is 34.9 *pm* and is less than the Bohr radius. This indicates that a photon can ionize a hydrogen atom at ground state only if its radius is less than the Bohr radius.

## IV. PHOTON TRAIN

### A. Copy of a photon and Fock state

An atom jumps from the ground state |a> to the excited sate |a'> by absorption of a photon. The excited atom returns to the ground state under stimulation of the photon $|1>_0$ and simultaneously gives off another photon $|1>_1$ which is the same as $|1>_0$ except for the phase according to the theory and experiment of stimulated emission. So, a photon and its copy constitute a phase-entangled photon pair. The excited atom |a'> also returns to the ground state under stimulation of the photon pair and simultaneously gives off another photon $|1>_2$ which is the same as both $|1>_0$ and $|1>_1$ except for the phase. As such, a phase-entangled n-photon train occurs by the sequential stimulated emissions and corresponds to the wave train in optics, and its electric field can be written as

$$E_n = f_n E_0 Re(e^{i\varphi} u), \qquad (21)$$

$$f_n = \begin{cases} 1, \rho < \rho_0 \text{ and } \varphi \in [\varphi_n, \varphi_0] \\ 0, \rho > \rho_0 \text{ or } \varphi \notin [\varphi_n, \varphi_0] \end{cases}, \qquad (22)$$

where $\varphi_n = \varphi_0 - n\pi$. The expression (21) is the same as (1) except for the phase interval. It follows that the phase-entangled n-photon train has the phase interval of $n\pi$ and n photons which are the same except for the phase, in which there is the phase difference of $\pi$ between the adjacent photons and the state vectors of different photons are all orthogonal each other, and can be described by the normalized state vector

$$|n\rangle = \frac{1}{\sqrt{n}} \sum_{j=0}^{n-1} |1\rangle_j. \qquad (23)$$

Since the number $n$ of photons in $|n\rangle$ has the definite value, $|n\rangle$ is an eigenstate of the number operator, and the phase-entangled n-photon train belongs to the number state or Fock state. A photon and its copy have the phase difference of $\pi$ and can be described by

$$|2\rangle = \frac{1}{\sqrt{2}} (|1\rangle_0 + |1\rangle_1). \qquad (24)$$

The electric fields of a linearly-polarized photon and its copy have the opposite directions.

### B. A laser beam and coherent state

A light beam is an ensemble of the phase-entangled n-photon trains with different wave vectors and can be described by the state vector

$$|\{n_{k_i}\}\rangle = \sum_{i=1}^{\infty} |n_{k_i}\rangle, \qquad (25)$$

where $n_{k_i}$ is the photon number in the i-th photon train with the wave vector $k_i$. The wave vectors are almost the same in a laser beam and $n_{k_i}$ can be simplified as $n$. The photon-counting experiments have showed that the photons from a laser well above the threshold obeys the Poisson distribution [22,23] and a laser beam can be described by the state vector

$$|\alpha\rangle = e^{-\bar{n}/2} \sum_{n=0}^{\infty} \frac{\alpha^n}{\sqrt{n!}} |n\rangle, \qquad (26)$$

where $\bar{n} = |\alpha|^2$ is the averaged value of $n$. It is follows that a laser beam is an ensemble of the phase-entangled n-photon trains and belongs to the coherent state, and the probability finding the n-photon train obeys the Poisson distribution, i.e.

$$P(n) = |\langle n|\alpha\rangle|^2 = \bar{n}^n e^{-\bar{n}}/n!. \qquad (27)$$

Since $\Delta n/\bar{n} = 1/\sqrt{\bar{n}}$ and $\bar{n}$ approaches infinity in a laser beam, the relative deviation is very close to zero, and the power and intensity of a laser beam are very stable.

The average number of the n-photon trains in any cross section of a laser beam is given by $\bar{m} = P/P_1$ where $P$ is the beam power and $P_1 = 2h\nu^2$ is power of the n-photon train. The intensity of the laser beam with the radius $r_0$ can be given by

$$I = \bar{m}(\rho_0/r_0)^2 I_1. \qquad (28)$$

where $I_1 = \varepsilon_0 E_0^2 c/2$ is the averaged intensity of the n-photon train. Since the beam radius is generally much larger than the photon radius, i.e. $r_0 >> \rho_0$, and the beam intensity is much weaker than the intensity of the n-photon train. A laser beam is usually described by the plane wave with the intensity $I = \varepsilon_0 E_p^2 c/2$ where

$E_p$ is amplitude of the electric field and satisfies the relation

$$E_p/E_0 = \sqrt{\bar{m}}\, \rho_0/r_0. \qquad (29)$$

For the 620-nm laser beam of $r_0$=0.091 mm, $\rho_0/r_0$=$10^{-6}$, $E_p/E_0$=$10^{-4}$ and $I/I_1$=$10^{-8}$ for $\bar{m}$=$10^4$. It follows that the electric field of a laser beam does not correspond to the true one at any position and is much weaker than electric field in the n-photon train. For $P < P_1$ and $\bar{m} < 1$, $\bar{n}$ becomes smaller and smaller with decrease of $P$, and the output power and intensity become more and more unstable. So, $P_1$ can be considered as the threshold power of a laser.

## V. SUMMARY

The present expressions of electromagnetic fields of a photon properly describe the particle characteristics of a photon. The length of a photon is half of the wavelength and the radius is proportional to square root of the wavelength. A photon can ionize a hydrogen atom at ground state only if its radius is less than the Bohr radius. A photon and its copy have the phase difference of $\pi$ and constitute a phase-entangled photon pair. The phase-entangled n-photon train results from the sequential stimulated emissions and belongs to the Fock state. A laser beam is an ensemble of the phase-entangled n-photon trains and belongs to the coherent state. The intensity of a laser beam is the averaged intensity of the n-photon trains and is much weaker than the intensity of the n-photon train. The threshold power of a laser is equal to the power of the n-photon train.

## ACKNOWLEDGMENTS

This work was supported by National Natural Science Foundation of China (No. 60778017)